\documentclass[aps,prd,showpacs,nofootinbib,floats,floatfix,preprintnumbers,groupedaddress,twocolumn]{revtex4}

\usepackage{bm}
\usepackage{latexsym}
\usepackage{dcolumn}
\usepackage{amsmath,amsfonts,amssymb}
\usepackage{graphicx,epsfig}
\usepackage{amsmath}
\usepackage{fancyhdr}
\usepackage{hyperref}
\usepackage{graphicx,epstopdf}
\hypersetup{
	colorlinks   = true, 
	urlcolor     = blue, 
	linkcolor    = blue, 
	citecolor   = red 
}

\begin{document}
\title{P-V criticality of AdS black holes in a general framework}
\author{Bibhas Ranjan Majhi$^{a}$\footnote {\color{blue} bibhas.majhi@iitg.ernet.in}}
\author{Saurav Samanta$^{b,c}$\footnote {\color{blue} srvsmnt@gmail.com}}

\affiliation{$^a$Department of Physics, Indian Institute of Technology Guwahati, Guwahati 781039, Assam, India\\
$^b$Department of Physics, Narasinha Dutt College, 129, Belilious Road, Howrah 711101, India\\
$^c$Department of Physics, Bajkul Milani Mahavidyalaya, P. O. - Kismat Bajkul, Dist. - Purba Medinipur, Pin - 721655, India
}

\date{\today}

\begin{abstract}
In black hole thermodynamics, it has been observed that AdS black holes behave as van der Waals system if one interprets the cosmological constant as a pressure term.  Also the critical exponents for the phase transition of AdS black holes and the van der Waals systems are same. Till now this type of analysis is done by two steps. In the first step one shows that a particular metric allows phase transition and in the second step, using this information, one calculates the exponents. Here, we present a different approach based on two universal inputs (the general forms of the Smarr formula and the first law of thermodynamics) and one assumption regarding the existence of van der Waal like critical point for a metric. We find that the same values of the critical exponents can be obtained by this approach. Thus we demonstrate that, though the existence of van der Waal like phase transition depends on specific metrics, the values of critical exponents are then fixed for that set of metrics. \end{abstract}


\maketitle

{\section{Introduction}}
   
   In conventional thermodynamics, phase transition is an important subject -- it explains how a substance undergoes a change from one phase to another. For black holes, the investigation of phase transition was started long ago by Davis \cite{Davies:1978mf} and was taken up later in a more rigorous way by Hawking and Page \cite{Hawking:1982dh}. Over the years, many different ways have been proposed to characterize and explore the different phases of a black hole. Recently two approaches attracted a lot of attention. In one approach, one mainly looks for the divergence of specific heat and inverse of isothermal compressibility \cite{Banerjee1}--\cite{Lala:2012jp}. Here black hole mass is treated as internal energy and cosmological constant ($\Lambda$), if it appears, is taken as a constant number. In the other approach, which is confined to AdS black holes, mass is treated as enthalpy and $\Lambda$ is regarded as a pressure term \cite{Dolan:2010ha,Dolan:2011xt}.
   An interesting feature of the second approach is that the thermodynamical variables of the black hole satisfy a van der Waal like equation of state and the critical exponents are identical to those of standard van der Waals system. This was first shown for charged $AdS$ black hole \cite{Kubiznak:2012wp} and then for higher dimensional charged and rotating black hole \cite{Gunasekaran:2012dq}. Later a similar study was done for charged topological black hole \cite{Zhao:2013oza}, quantum corrected black hole \cite{Ma:2014vxa}, dilaton black hole \cite{Dehghani:2014caa,xo}, Gauss-Bonnet black hole \cite{Cai:2013qga,Zou:2014mha} and also for Lovelock gravity \cite{Dolan:2014vba}-\cite{Xu:2014tja}, quasitopological gravity \cite{Mann} and conformal gravity \cite{Xu:2014kwa}. This extended phase space approach is also found to be quite successful in equilibrium state space geometry \cite{Hendi:2015hoa,Hendi:2016njy}. 
   
   
   A remarkable feature of these studies \cite{Kubiznak:2012wp}-\cite{xo} and similar other studies \cite{Zou:2013owa}-\cite{Azreg-Ainou:2014twa} (for review, see \cite{Altamirano:2014tva,Kubiznak:2016qmn}) is that the values of the critical exponents are independent of the metric if one restricts oneself to any one of the two approaches. Thus there is a clear signature of universality with the universality class being characterized by the approach one choses to use. It is only too natural to ask: what makes them universal? 
   
    Very recently we provided an method of exploring the critical exponents in the context of the first approach with a minimal amount of information about the metric  \cite{Mandal:2016anc}. Here black hole satisfies usual first law of thermodynamics, its mass is interpreted as internal energy and $\Lambda$ is treated as a pure constant having no significant role to play. We demonstrated how it is possible to extract the critical exponents with just a single assumption -- a phase transition point exists. It turned out that, this assumption makes the whole analysis very easy and transparent. In this paper our goal is the same but this time our study is addressed to the second approach. Here, we want to treat black hole mass as enthalpy so that $\Lambda$ can be treated as an intensive variable (i.e., a pressure-like term). 
    Reassuringly, using this assumption, we have been successful in calculating the critical exponents elegantly. This satisfactorily complements our previous work \cite{Mandal:2016anc}.  
    
  Let us now mention the things we achieved here. Present analysis clearly shows that if there is a van der Waal type phase transition for a set of black holes, the values of critical exponents must be same for that set and these values can be calculated without any further assumption about the spacetime. Although people intuitively expects this result, it has not been shown explicitly earlier. As already mentioned, before the calculation of critical exponents researchers give significant effort to find a spacetime which exhibits liquid-vapour type of phase transition. Our work shows that once this type of spacetime is found, calculation of critical exponents for that spacetime is not really necessary. In this sense, present paper nicely complement the existing literature.

   Before going into the main analysis, let us mention the essential inputs that have been used to achieve our goal. Since the general structure of the Smarr formula and the first law of thermodynamics are very much universal, they are considered to be valid in all black holes. In addition to this we also assume that there is a van der Waal like critical point about which two phases of black holes exist. These simple ingredients turn out to be adequate for our purpose. 
   
   {\it Notations}: $C_V$ and $K_T$ denote the specific heat at constant volume ($V$) and isothermal compressibility at constant temperature ($T$), respectively. $P$ stands for the pressure of the system. The critical values of pressure, temperature and volume are specified by $P_c$, $T_c$ and $V_c$. We mark order parameter, the difference between the volumes of two phases, by $\eta$.

\vskip 2mm
\noindent
{\section{Critical Phenomena: a unified picture}}
  The critical exponents ($\alpha, \beta, \gamma, \delta$) for a van der Waals thermodynamic system are defined as \cite{book}
\begin{eqnarray}
&&C_v \sim |t|^{-\alpha}~;
\nonumber
\\
&& \eta \sim v_l-v_s\sim |t|^{\beta}~;
\nonumber
\\
&& K_T = -\frac{1}{v}\Big(\frac{\partial v}{\partial P}\Big)_T \sim |t|^{-\gamma}~;
\nonumber
\\
&&P-P_c \sim |v-v_c|^{\delta}~;
\label{exponent}
\end{eqnarray}
where $t=T/T_c-1$. Here $v$ stands for specific volume which for a gaseous system is defined as volume per molecules of the system \cite{book}. Usually, every quantity is calculated in terms of the specific volume. $l$ and $s$ in the subscript refer to the two phases of the system; e.g. in the case of normal thermodynamics, $l$ and $s$ may be vapour and liquid phases.  
The critical point is a point of inflection, denoted by two conditions $(\partial P/\partial v)_T = 0 = (\partial^2P/\partial v^2)_T$ where $P$ is expressed as a function of temperature and specific volume. 
For the AdS black hole case, it has been observed that the specific volume is a function of $r_+$, the location of event horizon (e.g. see \cite{Kubiznak:2012wp} for AdS-RN black hole and for other cases, refs. can be availed from \cite{Altamirano:2014tva,Kubiznak:2016qmn}). On the other hand, as we shall see later, thermodynamic volume $V$, in general, is also a function of $r_+$ only {\footnote{In general, the thermodynamic volume ($V$) is a function of $r_+$, angular momentum and charges of the black hole \cite{Cvetic:2010jb}. We shall explain later that the whole analysis for Van der Waals like phase transition happens for fixed values of angular momentum and charges and so, in principle, we can take $V$ as a function of $r_+$ only.}}.  Therefore it is obvious that $P$ can also be expressed as function of $T$ and $V$. Now to find the critical exponent $\alpha$ one needs to find the entropy. This is done by taking the derivative of Helmholtz free energy (expressed in terms of $T$ and $r_+$), with respect to $T$, keeping $r_+$ constant. On the other hand, for other exponents, $P=P(T,v)$ has to be expanded around the critical point. One can verify that the expansion for $P (T,v)$ will have the same form as that for $P(T,V)$ where the variable $(v/v_c)-1$ plays the same role as $V/V_c - 1$. Hence, to find the critical exponents, we can work with $V$ instead of $v$ as well. So in our analysis, at the critical point, we impose the conditions  as
\begin{equation}
\frac{\partial P}{\partial V}=0 = \frac{\partial^2P}{\partial V^2}~.
\end{equation}
Here in taking the derivatives, $T$ is kept constant and $P$ is expressed as a function of $V$ and $T$. The same was also mentioned and adopted in the original work \cite{Kubiznak:2012wp} (see footnote $5$ of this paper). 
We shall use these conditions together with the basic definitions to find the values of the critical exponents for a AdS black hole.

  To start with, let us consider the two general results which are universally established: one is the general form of Smarr forumla and the other one is the first law of thermodynamics.
  The Smarr formula for a $D$-dimensional AdS black hole, in general, can be taken as
\begin{equation}
M=f_1(D)TS-f_2(D)PV + f_3(D)XY~,
\label{Smarr}
\end{equation}
where $M$, $T$ and $S$ are mass, temperature and entropy of the horizon while $P$ and $V$ are thermodynamic pressure and volume, respectively. The values of the unknown functions $f_1, f_2, f_3$ depend on the spacetime dimensions only. For example, in $D=4$, one has $f_1=2$, $f_2=2$ and $f_3=1$. We shall see that the explicit forms of them are not required for our main purpose. In the above equation, $X$ stands for electric potential $\Phi$ or angular velocity $\Omega$ or both whereas $Y$ refers to charge $Q$ or angular momentum $J$ or both for a black hole. The corresponding first law of thermodynamics is given by
\begin{equation}
dM = TdS+XdY+VdP~.
\label{law1}
\end{equation}
Note that, in the above expression $M$ is not the energy and the present form is not identical to the standard form of first law\footnote{Discussion on ``non--standard'' first law of thermodynamics and Smarr formula due to $\Lambda$ can be found in  \cite{Dolan:2010ha}, \cite{Dolan:2011xt}, \cite{Caldarelli:1999xj}-\cite{Dolan:2011jm}.}. But this can be expressed in the standard form:
\begin{equation}
dE=TdS+XdY-PdV 
\end{equation}
if one identifies the energy of the black hole as
\begin{equation}
E=M-PV~.
\label{energy}
\end{equation}
Thus $M$ is the enthalpy of black hole. Using (\ref{Smarr}), the Helmohtz free energy ($F$) of the system is found to be
\begin{equation}
F=E-TS=(f_1-1)TS - (f_2+1)PV+f_3XY~.
\label{Hfree}
\end{equation}

   Now for a black hole, if one keeps $Y$ constant then mass ($M$) has to be a function of pressure ($P$) and location of the horizon ($r_+$); i.e. $M=M(r_+,P)$. The reason is as follows. The location of the horizon is defined by the vanishing of the metric coefficient and in general this coefficient is a function of $M$, $Y$ and $P$ (as $P=-\Lambda/8\pi$) for a AdS black hole. Therefore, $r_+=r_+(M,Y,P)$ and hence for a fixed value of $Y$, $M$ is a function of $r_+$ and $P$. On the other hand, we have  $V=V(r_+)$ and in general $X=X(r_+)$. Here it must be pointed out that, in general $V$ depends on not only $r_+$ but also on the angular momentum or the existing charges of the black hole; i.e. on $Y$ (For example, see \cite{Cvetic:2010jb}). But since the analysis is done for fixed value of $Y$, without any loss of generality, we can take $V$ as function of $r_+$ only. Of course, if there is any deviation from it, this has to be treated separately. The first law of thermodynamics (\ref{law1}) shows that $S$ can be expressed as a function of the black hole parameters; i.e. $S = S(M,Y,P)$. So by the above argument we have $S=S(r_+,P)$. Therefore the Smarr formula (\ref{Smarr}) shows that for a fixed value of $Y$, one finds the solution for pressure from the equation
   \begin{equation}
    P = \frac{f_1TS(r_+,P)}{f_2V(r_+)}-\frac{M(r_+,P)}{f_2V(r_+)}+\frac{f_3X(r_+)Y}{f_2V(r_+)}~,
    \label{P}
   \end{equation}
which has to be function of both temperature and horizon radius; i.e. $P=P(T,r_+)$. Hence the free energy (\ref{Hfree}) as well as $S$ are in general  functions of both $T$ and $r_+$. Since entropy can be calculated from the relation $S=-(\partial F/\partial T)_V$ ($\equiv$ $-(\partial F/\partial T)_{r_+}$ as $V=V(r_+)$), the evaluated value can be, in general, function of both temperature and horizon radius.  Now we find from (\ref{Hfree})
\begin{eqnarray}
-S=\Big(\frac{\partial F}{\partial T}\Big)_{r_+} &=& (f_1-1)S + (f_1-1) T \Big(\frac{\partial S}{\partial T}\Big)_{r_+}
\nonumber
\\
&-&(f_2+1)V(r_+)\Big(\frac{\partial P}{\partial T}\Big)_{r_+}~.
\label{PF}
\end{eqnarray}
This gives,
\begin{eqnarray}
0 &=& f_1S + (f_1-1) T \Big(\frac{\partial S}{\partial T}\Big)_{r_+}
\nonumber
\\
&-&(f_2+1)V(r_+)\Big(\frac{\partial P}{\partial T}\Big)_{r_+}~.
\label{PF1}
\end{eqnarray}

For a large class of black holes, it is found that pressure is a linear function of temperature, i.e.
\begin{eqnarray}
P=p_0(r_+)+Tp_1(r_+)~.
\label{PTR}
\end{eqnarray}
This, we want to take as an input for the AdS black holes because it has been observed that those which exhibit van der Waals like behaviour, have the above feature when $P$ is expressed as a function of $T$ and $r_+$. The same also happens if one looks at the van der Waals equation of state (e.g. see \cite{book,Kubiznak:2012wp}). Another way to see it is as follows. In later discussion we shall see that near the critical point, $P$ to the leading order, is a linear function of $t=(T/T_c)-1$. Since all the critical exponents are determined by the analysis very near to the critical point, there is no loss of generality to take an expression like (\ref{PTR}). 
This, when substituted in (\ref{PF1}), gives
\begin{eqnarray}
f_1S+ (f_1-1) T \Big(\frac{\partial S}{\partial T}\Big)_{r_+}-f_1a(r_+)=0
\end{eqnarray}
where $a(r_+)=\frac{1}{f_1}(f_2+1)V(r_+)p_1(r_+)$. The solution of above first order differential equation is
\begin{eqnarray}
S=a(r_+)+C(r_+)T^{-\frac{f_1}{f_1-1}}
\end{eqnarray}
where $C(r_+)$ is an integration constant. Since $f_1/(f_1-1)$ is always positive for $D\geq 4$, the last term of the R.H.S. diverges as $T\rightarrow 0$. So to be consistent with the third law of thermodynamics, $C(r_+)$ must be zero. This leads to $(\partial S/\partial T)_V =(\partial S/\partial T)_{r_+}= 0$;
i.e. the specific heat at constant volume $C_V = T(\partial S/\partial T)_{V} = 0$ and hence by (\ref{exponent}) one finds the value of the critical exponent $\alpha=0$.

  Next we shall find the other critical exponents. Remember that the pressure here is in general a function of both $T$ and $r_+$. Now since thermodynamic volume $V$ is a function of $r_+$ only, we take $P$ as function of $T$ and $V$ for our purpose. Let us expand $P(T,V)$ around the critical values $T_c$ and $V_c$:
  \begin{eqnarray}
  P = &&P_c + \Big[\Big(\frac{\partial P}{\partial T}\Big)_{V}\Big]_c (T-T_c)
  \nonumber
  \\
  &&+ \frac{1}{2!} \Big[\Big(\frac{\partial^2 P}{\partial^2 T}\Big)_{V}\Big]_c (T-T_c)^2
  \nonumber
  \\
  &&+ \Big[\Big(\frac{\partial^2P}{\partial T\partial V}\Big)\Big]_c(T-T_c)(V-V_c)
  \nonumber
  \\
  &&+\frac{1}{3!}\Big[\Big(\frac{\partial^3P}{\partial V^3}\Big)_T\Big]_c (V-V_c)^3+\dots~.
  \end{eqnarray}
In the above expression, we have used the fact that at the critical point $(\partial P/\partial V)_c=0=(\partial^2P/\partial V^2)_c$. Defining two new variables $t=T/T_c-1$ and $\omega = V/V_c-1$, and above expression is written as,\begin{equation}
P=P_c+Rt+Bt\omega+D\omega^3+Kt^2;
\label{Pexpansion}
\end{equation}
where $R, B, D,K$, etc. are constants calculated from the derivatives at the critical point. Here we have ignored the other higher order terms since they are very small.

   Now for the usual van der Waal's system, below critical point there is a portion between the vapor and liquid phases where both of them coexists which is either unstable or meta-stable and therefore experimentally this exact form of the isotherm is not observed. Hence the ends of the vapor and liquid phases are usually connected by a straight line, parallel to the volume axis, to match with the obtained isotherm for a substance by direct experiment. Then one obtains two portions of van der Waal's isotherm and the straight line is drawn in a such a way that the area of both of them are equal so that one has $\oint VdP = 0$. This is known as Maxwell's area law. Since we are treating the AdS black holes similar to van der Waal's system, this law can be used here also \cite{Wei:2014qwa,Belhaj:2014eha}. In addition, by keeping the analogy between the volumes of vapor and liquid phases, it will be denoted that $\omega_l$ and $\omega_s$ correspond to the large and small volumes of black hole in two different phases. In between these two we have mixture of both of them.
   Since we are interested in the isotherms, to use Maxwell's area law we find $dP = (Bt+3D\omega^2)d\omega$ for constant $t$, which yields
\begin{equation}
\int_{\omega_l}^{\omega_s}\omega(Bt+3D\omega^2)d\omega + \int_{\omega_l}^{\omega_s}(Bt+3D\omega^2)d\omega =0~.
\label{omega}
\end{equation}
Next note that for the usual system, the end point of vapor and the staring point of liquid have same pressure. Similarly, here also the pressure does not change, i.e. $P_l=P_s$, and then (\ref{Pexpansion}) implies 
\begin{equation}
Bt(\omega_l-\omega_s) + D(\omega_l^3-\omega_s^3)=0~.
\label{first}
\end{equation}
So the second integral of (\ref{omega}) vanishes. Therefore it reduces to
\begin{equation}
Bt(\omega_l^2-\omega_s^2)+\frac{3D}{2}(\omega_l^4-\omega_s^4) = 0~.
\label{second}
\end{equation}
One can now easily find the non-trivial solutions of the above two equations. These are $\omega_l =(-Bt/D)^{1/2}$ and $\omega_s=-(-Bt/D)^{1/2}$. Therefore we find 
\begin{equation}
\eta \sim V_l-V_s = (\omega_l-\omega_s)V_c\sim |t|^{1/2}~, 
\end{equation}
which yields $\beta=1/2$.

     To find $\gamma$ we need to calculate $K_T$, given in (\ref{exponent}). So we first find $(\partial P/\partial V)_T$ from (\ref{Pexpansion}). Upto first (leading) order it is given by
     \begin{equation}
     \Big(\frac{\partial P}{\partial V}\Big)_T \simeq \frac{B}{V_c}t~,
     \end{equation}
where one needed to use $\partial \omega/\partial V = 1/V_c$. Therefore the value of $K_T$ near the critical point is 
\begin{equation}
K_T \simeq \frac{1}{Bt}\sim t^{-1}~.
\end{equation}
This implies $\gamma=1$. Next for $T=T_c$, the expression (\ref{Pexpansion}) for pressure yields
\begin{equation}
P-P_c\sim \omega^3\sim (V-V_c)^3~; 
\end{equation}
i.e. the value of the critical exponent is $\delta=3$.
It must be pointed out that the derived critical exponents satisfy the following scaling laws:
\begin{eqnarray}
&&\alpha+2\beta+\gamma=2; 
\nonumber
\\
&&\gamma=\beta(\delta-1)~.
\label{scaling}
\end{eqnarray}
As is well known, these scaling laws are universal in nature. Their thermodynamic analysis can be found in \cite{Stanley1}.

    Note that in our analysis we consider the thermodynamic volume as function of horizon radius only. In general, this is not always true. For example when black hole has some hair, $V$ is not just the volume inside the black hole but also the integral of the scalar potential and hence it depends on both $r_+$ and the charges corresponding to the hairs (for instance, see \cite{Cvetic:2010jb,Caceres:2015vsa}). Also for some black hole solutions, it may be possible that the entropy is not only a function of $r_+$ and $T$ but also function of other parameters of the spacetime.  There is a hairy case \cite{Hennigar:2015wxa} in which this happens though volume is function of $r_+$ alone. Then it might come in mind that the present general approach incorporates only those cases which are consistent with the {\it no hair theorem}. In this regard, remember that in our present analysis, $Y$ is kept constant; i.e. we are taking the canonical ensemble. Then if there is any hairy charge that should not be varied also and hence $V$ will be a function of $r_+$ alone. In this situation, such solutions behaves as Van der-Waal's system. Of course for an analysis in the grand canonical ensemble, one needs to be careful and must deal them in a separate way.
There is another example \cite{Sadeghi:2016dvc} where $S=S(r_+,T)$ corresponding to a solution with logarithmic correction to entropy. This can be dealt in this present approach.

\vskip 2mm
\noindent
{\section{Conclusions}}	
Using the tools of thermodynamics, phase transition of completely dissimilar systems like chemical, magnetic, hydrodynamic etc. has been studied thoroughly. Black hole phase transition is a relatively new observation which deserves careful study. Though the first order phase transition from non-extremal to extremal black hole is known for some time, there are various other types of phase transition. A new type of phase transition has been found recently where cosmological constant ($\Lambda$) is treated as dynamical variable (instead of a constant) equivalent to pressure of hydrodynamic system \cite{Kubiznak:2012wp}. In this interpretation, phase transition of different black holes has been shown to be quite analogous to van der Waal's system. Interestingly critical exponents found from different metrics are same. This naturally deserves some explanation.

It is well known that critical exponents of quite different systems can be same. But for black holes, this point is not well appreciated. In our previous work \cite{Mandal:2016anc} we showed that, starting from few very general assumptions about the black hole spacetime, critical exponents can be calculated. In that work we did not treat $\Lambda$ as a dynamical variable. Present paper is a continuation of our previous work. Here we take a different interpretation of $\Lambda$. In this work we showed that, the single assumption of existence of liquid-vapour type phase transition for a spacetime determines the values of critical exponents. No other assumption regarding the variables or dimensions of spacetime is necessary.

Till now values of critical exponents are obtained first by finding a suitable metric which exhibits this type of phase transition and then by explicit calculation using the definitions of exponents. Though expected, it remained unexplained why these values are same for different $AdS$ black holes. In this paper, we give a completely satisfactory answer to this question. Thus our present work nicely complements the existing scientific works in this field.  

It may be mentioned that, the analysis shows that the values of the critical exponents are very much universal in nature and also they are independent of spacetime dimensions. In literature, different critical exponents have been found very recently \cite{Mann}, usually known as ``nonstandard'' critical exponents, for black hole solutions in higher curvature gravity theory. These are not due to the analysis around the standard critical point; rather around {\it isolated} one. In that respect our analysis is different from them. Of course it would be very much worthwhile to look at such non-standard phase transition. 
    
\vskip 4mm
{\section*{Acknowledgments}}
We thank to the anonymous referees for pointing out some important issues which helped to improve the earlier version. We also thank Juan F. Pedraza for making several interesting and valuable comments on the first version of our paper.
The research of one of the authors (BRM) is supported by a START-UP RESEARCH GRANT (No. SG/PHY/P/BRM/01) from Indian Institute of Technology
Guwahati, India. 

\end{document}